\documentclass[twocolumn,trackchanges]{aastex7}

\usepackage{xspace}
\usepackage{multirow}
\usepackage{natbib}
\usepackage{mathtools}
\usepackage{comment}

\newcommand{\be}{\begin{equation}}
\newcommand{\ee}{\end{equation}}
\newcommand{\bea}{\begin{eqnarray}}
\newcommand{\eea}{\end{eqnarray}}
\newcommand{\Msol}{\hbox{M$_\odot$}}
\newcommand{\Lsol}{\hbox{L$_\odot$}}

\newcommand{\jwst}{{JWST\null}\xspace}
\newcommand{\hst}{{HST\null}\xspace}
\defcitealias{harris_reinacampos2023} {Paper~1}

\begin{document}
\title{PEARLS: Globular Clusters and Ultra-Compact Dwarfs in the El Gordo Galaxies at \boldmath{$z=0.87$}}

\author[orcid=0000-0001-8762-5772,sname='Harris']{William E. Harris} 
\affiliation{Department of Physics \& Astronomy, McMaster University, Hamilton ON L8S 4M1, Canada}
\email[show]{harrisw@mcmaster.ca}

\author[orcid=0000-0002-8556-4280,sname='Reina-Campos']{Marta Reina-Campos}
\affiliation{Canadian Institute for Theoretical Astrophysics (CITA), University of Toronto, 60 St George St, Toronto, M5S 3H8, Canada}
\affiliation{Department of Physics \& Astronomy, McMaster University,  Hamilton ON L8S 4M1, Canada}
\email[show]{reinacampos@cita.utoronto.ca}

\author[orcid=0000-0002-6610-2048]{Anton M. Koekemoer}
\affiliation{Space Telescope Science Institute, 3700 San Martin Drive,Baltimore MD 21218 USA}
\email{koekemoer@stsci.edu}

\author[0000-0001-6265-0541]{Jessica M.\ Berkheimer}
\affiliation{School of Earth and Space Exploration, Arizona State University, Tempe, AZ 85287, USA}
\email{jberkhei@asu.edu}

\author[0000-0001-6650-2853]{Timothy Carleton}
\affiliation{School of Earth and Space Exploration, Arizona State University, Tempe, AZ 85287, USA}
\email{tcarlet@asu.edu}

\author[0000-0003-3329-1337]{Seth H.\ Cohen}
\affiliation{School of Earth and Space Exploration, Arizona State University, Tempe, AZ 85287, USA}
\email{seth.cohen@asu.edu}

\author[0000-0003-1625-8009]{Brenda L.~Frye}
\affiliation{Department of Astronomy/Steward Observatory, University of Arizona, 933 N. Cherry Avenue, Tucson, AZ 85721, USA}
\email{brendafrye@gmail.com}

\author[0009-0008-0376-3771]{Tyler R.\ Hinrichs}
\affiliation{School of Earth and Space Exploration, Arizona State University, Tempe, AZ 85287, USA}
\email{trhinric@asu.edu}

\author[0000-0002-4884-6756]{Benne W.\ Holwerda}
\affiliation{University of Louisville, Department of Physics and Astronomy, 102 Natural Science Building, Louisville, KY  40292, USA}
\email{benne.holwerda@gmail.com}

\author[0000-0002-9984-4937]{Rachel Honor}
\affiliation{School of Earth and Space Exploration, Arizona State University, Tempe, AZ 85287, USA}
\email{rchonor@asu.edu}

\author[0000-0003-4223-7324]{Massimo Ricotti}
\affiliation{Department of Astronomy, University of Maryland, College Park, 20742, USA}
\email{ricotti@umd.edu}

\author[orcid=0000-0002-9895-5758,sname='Willner']{S.\ P.\ Willner}
\affiliation{Center for Astrophysics \textbar\ Harvard \& Smithsonian, 60 Garden Street, Cambridge MA 02138, USA}
\email{swillner@cfa.harvard.edu}

\author[0000-0001-8156-6281]{Rogier A.\ Windhorst} 
\affiliation{School of Earth and Space Exploration, Arizona State University, Tempe, AZ 85287, USA}
\email{Rogier.Windhorst@gmail.com}

\author[0000-0001-7592-7714]{Haojing Yan}
\affiliation{Department of Physics and Astronomy, University of Missouri, Columbia MO 65211, USA}
\email{yanha@missouri.edu}




\begin{abstract}

JWST/NIRCam 0.9 to 2.0~\micron\ images reveal a population of point sources around the major galaxies in the El Gordo cluster at redshift $z=0.87$.  
Their distribution in the color--magnitude diagrams shows a narrow sequence well separated from field-galaxy contamination and consistent with their identification as ultra-compact dwarf galaxies (UCDs) or luminous globular clusters (GCs).  
The point-source sequence is more luminous by almost a magnitude than the corresponding sequence in Abell~2744 at $z=0.31$, matching 
the predicted evolutionary change for GC/UCDs over the 4-Gyr difference in lookback time between these two clusters.  Deeper observations should allow direct JWST imaging of GC/UCD populations, even without the help of lensing, up to $z\sim 1.4$, a lookback time of more than 9~Gyr. Such observations would directly reveal the evolution of these compact stellar systems two-thirds of the way back to the Big Bang. 
\end{abstract}

\keywords{Globular star clusters; rich galaxy clusters; galaxy evolution; photometry; space telescopes}


\setcounter{footnote}{0}

\section{Introduction}

The James Webb Space Telescope (\jwst) is now enabling us to reconstruct the details of  early galaxy evolution with an entirely new range of observational material.
\textit{Lensing clusters} of galaxies at redshifts $z \gtrsim 0.2$ have been especially prominent targets, notably through the PEARLS \citep{windhorst+2023}, CANUCS \citep{willott+2022}, and UNCOVER \citep{bezanson+2024} NIRCam imaging surveys. Among the results made possible with this new material is the ability to survey the populations of globular clusters (GCs) and Ultra-Compact Dwarfs (UCDs) around the major galaxies in the target lensing clusters, extending to distances far beyond the range that could be reached by previous facilities and thus allowing us to see the state of those compact stellar systems systems many gigayears in the past.

The new \jwst imaging era for 
distant GCs and UCDs
has just begun. Early studies include those of \citet{faisst+2022,lee+2022,diego+2023,harris_reinacampos2023,harris_reinacampos2024,martis+2024}.
The two studies by Harris \& Reina-Campos (hereafter HR23, HR24)
analyzed deep \jwst photometry, derived from the UNCOVER database, of the GC systems in Abell~2744 (hereafter A2744) at $z=0.308$.  Adding targets at a wider range of redshifts will provide color--magnitude diagrams (CMDs) and luminosity functions (GCLFs) for GC populations extending over many Gyr of lookback time.  This capability is opening up a new way to test the predicted evolution of GC systems in the mass--metallicity plane \citep{choksi_gnedin2019,reina-campos+2022b,chen_gnedin2023}.  In particular, it should be  possible to track the luminosity evolution of the GC population produced by the combined effects of stellar evolution within GCs and dynamical evolution of GCs within the tidal field of their host galaxy.

\begin{figure*}
    \centering
    \includegraphics[width=\linewidth]{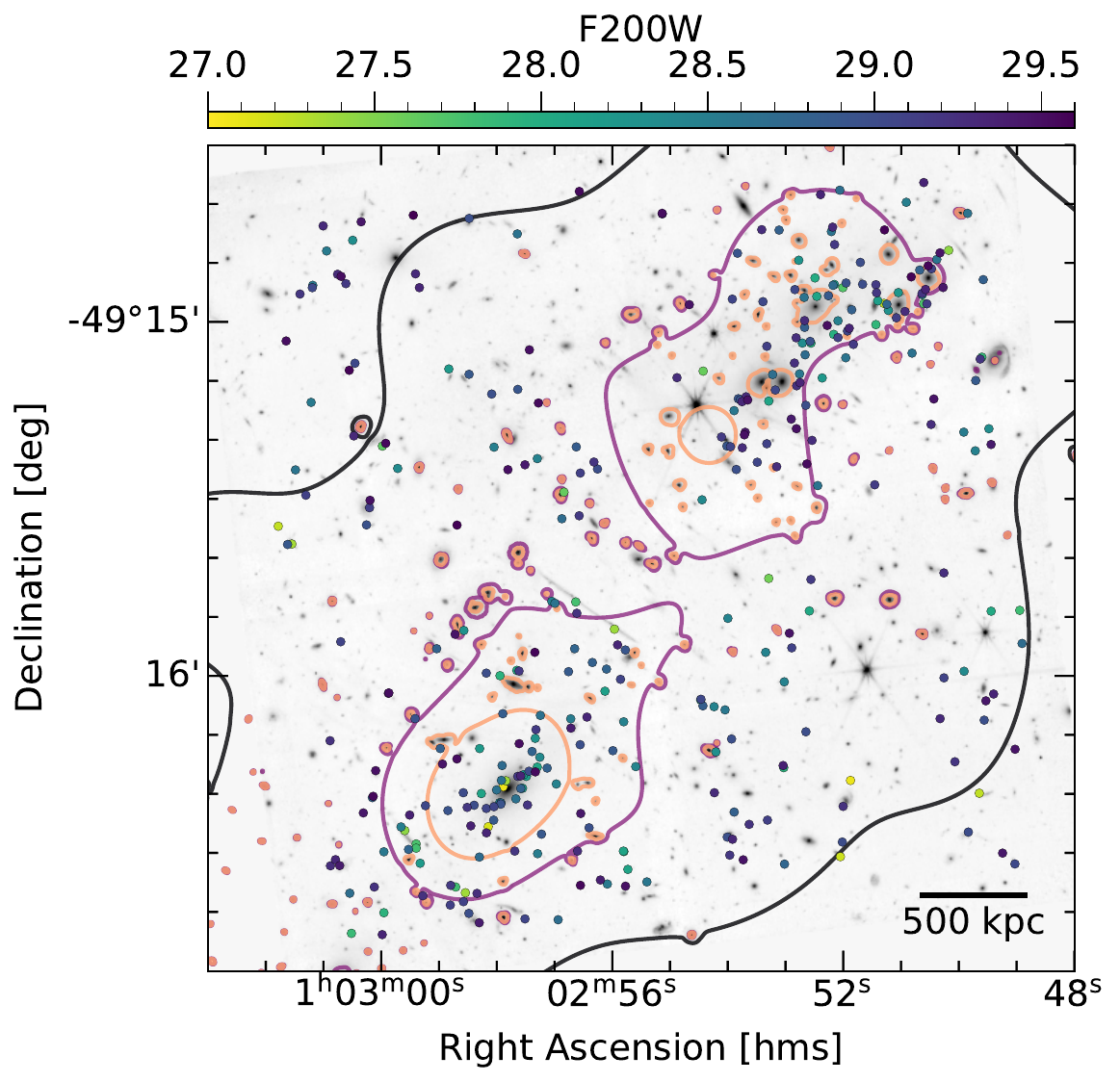}
    \caption{View of the El Gordo galaxy cluster. North is at top and East at left. The background is the NIRCam F200W image as a negative gray scale with logarithmic stretch. The contours represent the $50$th, $90$th, and $98$th percentiles of the mass distribution derived from strong lensing \citep{diego+2023}, and the filled circles mark the point sources in the field with $\rm F200W \le 29.6$.  The color of each circle indicates the brightness of the object according to the color scale at top.}
    \label{fig:field}
\end{figure*}

One of the most distant lensing clusters surveyed to date is El Gordo (=ACT-CL J0102$-$4915)
at redshift $z = 0.87$. A \jwst NIRCam image of the El Gordo field is shown in Figure~\ref{fig:field}. Although many large clusters of galaxies had already begun assembly at this early epoch, El Gordo has drawn particular interest because of its high total mass, which would make it exceptional even in the present-day Universe.   Several lines of evidence including gravitational lensing, X-ray intracluster gas, the Sunyaev--Zeldovich effect, and the velocity dispersion of El Gordo's member galaxies give a virial mass  up to $M_{\rm vir} \simeq 2 \times 10^{15}$~\Msol, corresponding to a virial radius $R_{\rm vir} \simeq 2.7$~Mpc.  Within the current $\Lambda$CDM paradigm, this high mass pushes the boundary of what should be the maximum mass that can be assembled only 6~Gyr after the Big Bang \citep{menanteau+2012,lindner+2014,kim+2021,asencio+2023,frye+2023,diego+2023}.

El Gordo has clear substructure and is in an active assembly stage with two major subcenters:  a region to the northwest with a handful of major galaxies and their many satellites, and a region of similar mass to the southeast with the brightest single member (the brightest cluster galaxy or BCG)\null.  This substructure is reflected in the spatial structure of the radio halo, the X-ray gas, and the dark-matter (DM) distribution inferred from gravitational lensing.  The clearest indication of this two-component structure may be in the lensing maps \citep{jee+2014,kim+2021,caminha+2023,frye+2023,diego+2023}, which show two major concentrations around the BCG in the SE quadrant and the giant galaxies in the NW quadrant.  An illustration of the contours from the lensing map of \citet{diego+2023} is in Fig.~\ref{fig:field}. Radio contours show two clear components as well, marking the direction of the collision between the two original sub-clusters \citep{lindner+2014,botteon+2016,kale+2025}. The X-ray maps \citep{botteon+2016,caminha+2023,frye+2023,diego+2023} paint a similar picture; \citet{zhang+2015} noted especially ``a remarkable wake-like structure'' trailing the SE component.  Modeling of the distribution \citep[e.g.,][]{molnar_broadhurst2015,ng+2015,zhang+2015} calls for an interpretation where we are now seeing the two original components having passed through each other at very high speed (twice the velocity dispersion) but not yet having fallen back to eventually virialize into a single structure.  Finally, the distribution of spectroscopically identified member galaxies \citep[][see these papers for contour maps]{caminha+2023,diego+2023,frye+2023}  generally follows the DM distribution.  The major galaxies themselves (the BCG in the SE and the handful of comparably large giants in the NW) show little evidence of any active star formation.

GCs and UCDs are also found prominently and in large numbers in BCGs, giant galaxies, and their host galaxy clusters.  Digital imaging and photometry of GC systems in large galaxies has a decades-long history, beginning with ground-based CCD imaging; a representative sampling of these studies in historical order includes \citet[][among others]{vandenbergh+1985,couture+1990,harris+1991,mclaughlin+1994,zepf_ashman1994,geisler+1996,harris+1998,lee+1998,rhode_zepf2004,forbes+2004,harris+2004,tamura+2006,bassino+2006,rhode+2007,bassino+2008,harris2009b,forbes+2011,faifer+2011,blom+2012,kartha+2014,durrell+2014,ennis+2019,ennis+2020,ennis+2024}.

Starting a decade later, \hst visible and near-IR cameras considerably extended the reach and depth of such studies. A partial list includes \citet[][]{whitmore+1995,woodworth_harris2000,kundu_whitmore2001,larsen+2001,peng+2006,jordan+2007,harris+2009,harris2009,villegas+2010,harris+2010,peng+2011,jennings+2014,forbes+2014,harris+2016,lee_jang2016b,cho+2016,alamo-martinez+2017,alamo-martinez+2021,harris2023,dornan_harris2023,hartman+2023} among others.  These studies provide a wealth of data for GC systems in the local Universe and form the basis for exploring GC systems at higher redshift and significant lookback times.

This paper reports a first test of the ability of \jwst to detect and resolve GC populations at redshifts approaching $z \sim 1$.  
Scaling from the GC observations in A2744 (HR23) and simply applying the $\sim$2.7-magnitude difference in distance modulus, one would expect the brightest GCs in  El Gordo galaxies to become detectable at $\rm F150W \sim 28$, which is well within reach of \jwst.  

The paper is organized as follows:
Section~2 describes the imaging data and the photometric measurement procedures.  Section~3 presents the resulting color--magnitude diagrams (CMDs) and a direct comparison with  previous results for A2744.
Section 4 summarizes the findings and discusses prospects for further work. 
We adopt the Planck 2015 cosmological parameters $H_0 = 67.8$ km s$^{-1}$ and $\Omega_{\Lambda} = 0.692$ \citep{planck2016}.  For El Gordo's $z = 0.87$, the luminosity distance $d_L = 5720$~Mpc for a distance modulus $(m-M)_0 = 43.79$.  The angular-size distance  $d_A = 1630$~Mpc gives an angular scale of 7.9~kpc~arcsec$^{-1}$. The lookback time is 7.36~Gyr, and the Universe's age was 6.4~Gyr.  In what follows, all magnitudes are on the AB system, and the notation `F200W' refers to either that JWST filter or to the AB magnitude measured through it, depending on context.  The foreground extinction (NED database) $A_V = 0.027$ has a negligible effect on the near-infrared magnitudes and colors used here.  

\section{Steps in the Photometry}


\subsection{Measurement Procedures}

Our present study of El Gordo uses NIRCam exposures taken as part of the PEARLS program \citep{windhorst+2023}.  The filters used here, all in the short-wave camera (SWC), and their exposure times were F090W (2491~s), F115W (2491~s), F150W (1890~s), and F200W (2104~s). These exposures were shorter than those for some other PEARLS target clusters but deep enough to allow a first test of the ability of \jwst to resolve GC populations in this redshift range.  The data are the same as used by \citet{frye+2023} and \citet{diego+2023} but from a new reduction using the \jwst pipeline version 11.17.2 in context jwst\_1100.pmap.
Inspection of these images, especially F150W and F200W, clearly shows some dozens of point sources in the right magnitude range to be GCs or UCDs grouped around the El Gordo major galaxies, and these are the targets for the present study. 

Methods for photometry of GC populations around large galaxies have been well developed over a long series of studies (references cited above).  The specific procedures used here, adapted for the \jwst NIRCam images, follow the steps  outlined in more detail by HR23, HR24.
Briefly, the measurement process started with mosaic images in each filter drizzled to a scale of 0\farcs03~pixel$^{-1}$, close to the native SWC pixel size \citep{windhorst+2023}.   All four filters were added to create a single deep stacked image that was then used to construct a finding list of objects for further detailed measurement. Because of gaps between the four individual detector elements in the SWC,\footnote{\url{https://jwst-docs.stsci.edu/jwst-near-infrared-camera/nircam-instrumentation/nircam-field-of-view}} the useful parts of the image are four squares $\sim$57\arcsec\ on a side separated by $\sim$10\arcsec. 
Long-wave-camera (LWC) images exist but are not deep enough to give useful information on the GCs.

Object detection and photometry was done with the  \texttt{DAOPHOT} code  \citep{stetson1987} in its \texttt{IRAF} implementation. Objects 4$\sigma$ above the local sky noise on the deep stacked image defined the finding list, which was then used for aperture photometry with \texttt{phot} in a 2.5-pixel aperture radius and then for final photometry via point-spread function (PSF) fitting with \texttt{allstar}.  Although bright, isolated foreground stars are rare in this high-latitude field, there are enough of them to define accurate empirical PSFs; eight such stars were used in each filter for this purpose. Correction of the \texttt{allstar} magnitudes to large radius was done using the curves of growth of aperture photometry out to 10 pixel radius for the bright PSF stars, extended to large radius with the encircled-energy curves supplied on the NIRCam/SWC webpages.\footnote{\url{https://jwst-docs.stsci.edu/jwst-near-infrared-camera/nircam-performance/nircam-point-spread-functions}} 

\begin{figure}
    \centering{
    \includegraphics[width=0.90\columnwidth]{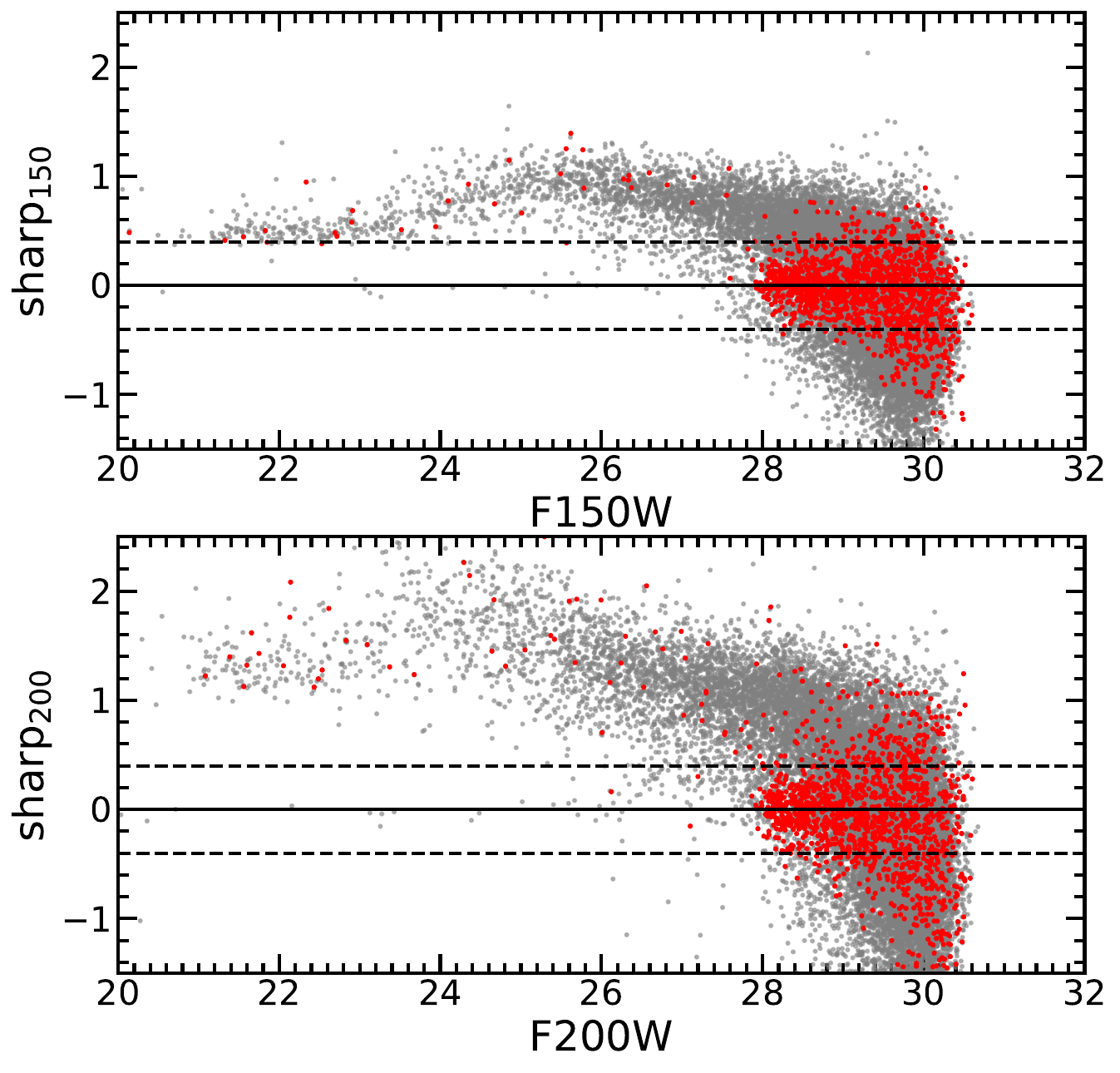}
  \caption{\texttt{DAOPHOT} parameter \texttt{sharp} plotted versus F150W (upper panel) and F200W (lower panel).  Grey symbols show all measured objects before culling.  Red symbols show the locations of the artificial stars.  Objects falling in the range $-0.4 \le sharp \le +0.4$ (dashed lines) were kept for the final list, and any falling outside those boundaries were rejected as nonstellar.]} 
  \label{fig:sharp}}
\end{figure}

\begin{figure}
    \centering{
    \includegraphics[width=0.95\columnwidth]{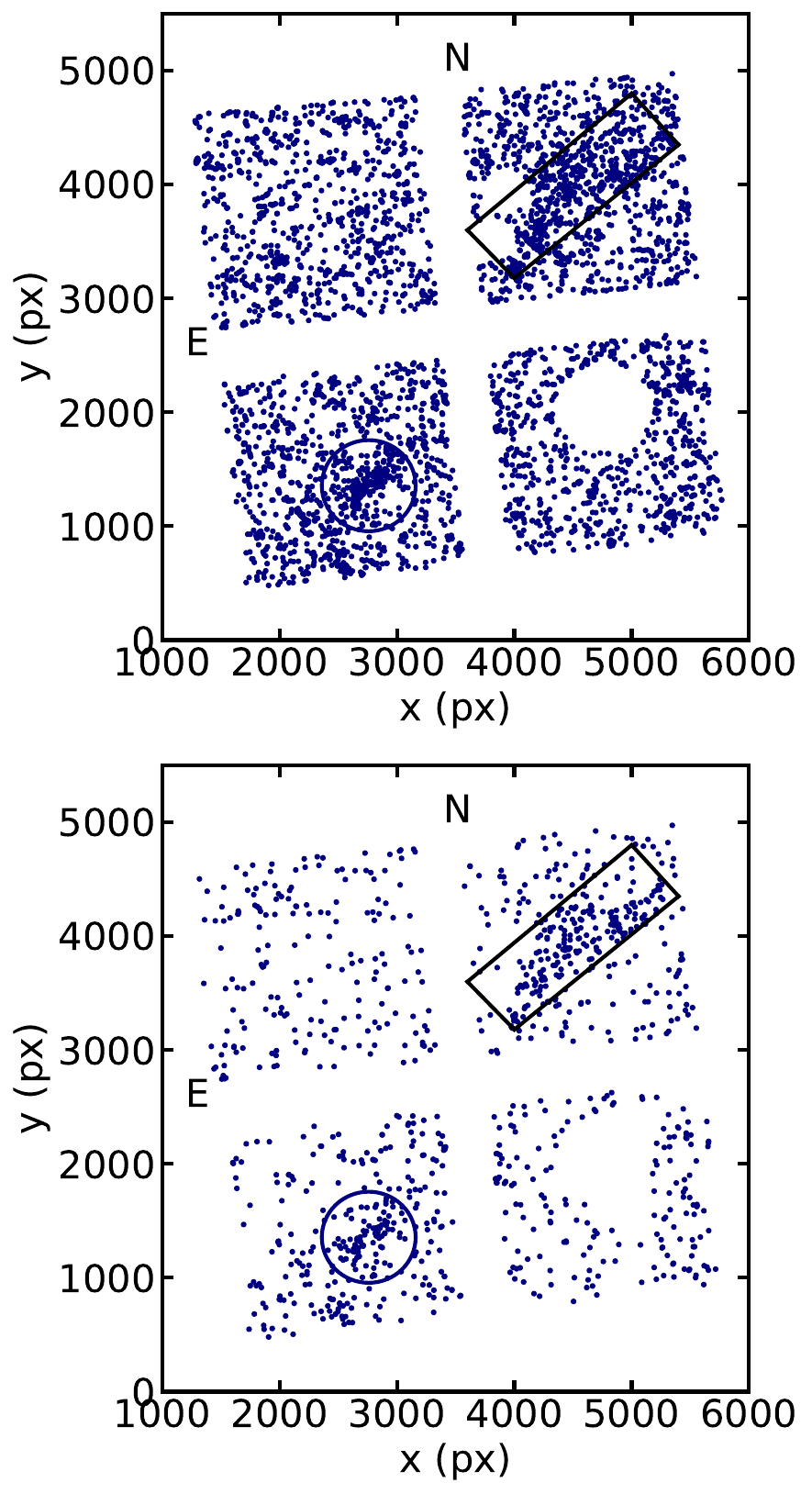}
  \caption{\emph{Upper panel:} Positions of all 4508 objects in the field.  The only culling step applied here was to remove points along the noisier detector margins and three circular areas affected by bright stars in the NW and SW quadtrants.  The solid circle  marked in the SE quadrant encloses the BCG, and the rectangle in the NW quadrant encloses the four brightest galaxies in that region. The areas inside the (circle+rectangle) together are referred to as ``the BCG region.'' \emph{Lower panel:} Positions of 1009 objects qualifying as point sources after culling by \texttt{sharp} as described in Section~\ref{s:cull}.} 
  \label{fig:xy2}}
\end{figure}

Artificial-star tests were used to estimate the photometric completeness levels of the measurements.  With the \texttt{addstar} function in \texttt{daophot}, $\simeq$2000 stars (scaled PSFs) covering the magnitude range 28.0--31.5 were added to the images at random positions.  The resulting images were then remeasured as described above.  Artificial stars that were successfully found and that also passed the culling step (see below)   
were classified as \emph{recovered} \citep[as defined by][]{harris_speagle2024}.
A  regression solution of the form
\begin{equation}
      p = \frac{1}{1+e^{-({\beta_0 + \beta_1 x_1)}}} ,
      \label{eq:simplelr}
\end{equation}
where $x_1 =\rm  F150W$ was then used to solve for the recovery probability $p$ as a function of F150W\null.  The result for the coefficients is $\beta_0 = 53.87 \pm 2.32, \beta_1 =-1.814 \pm 0.078$.  These give a 50\% recovery probability at F150W$_{0.5} = 29.7$ (the ``limiting magnitude''), with 80\% and 20\% levels at F150W$_{0.8} = 28.9$, and F150W$_{0.2} = 30.4$.  \citep[See][for a more extensive discussion of completeness modeling.]{harris_speagle2024}

\subsection{Selection and Culling}
\label{s:cull}
The \texttt{allstar} output parameter \texttt{sharp} was used to select point sources and help reject non-stellar  objects, which consist mostly of faint background galaxies too extended to match the PSF\null. Other artifacts include random noise spikes or cosmic-ray hits that are too narrow to match the PSF.  The adopted boundary for inclusion in the final list was abs(\texttt{sharp}) $ < 0.4$ in either of F150W or F200W, the two deepest images. These exclusion boundaries were based on the \texttt{sharp} values for artificial stars as illustrated in Figure~\ref{fig:sharp}.  The parameter \texttt{chi} can in principle also be used for culling, but in practice it was strongly correlated with \texttt{sharp} and did not yield many additional rejections.  Contamination from false detections was reduced further by including only objects that appeared and were matched on more than one filter.  Lastly, all candidate objects on the final list were confirmed by visual inspection.

\begin{figure}
    \centering{
    \includegraphics[width=0.99\columnwidth]{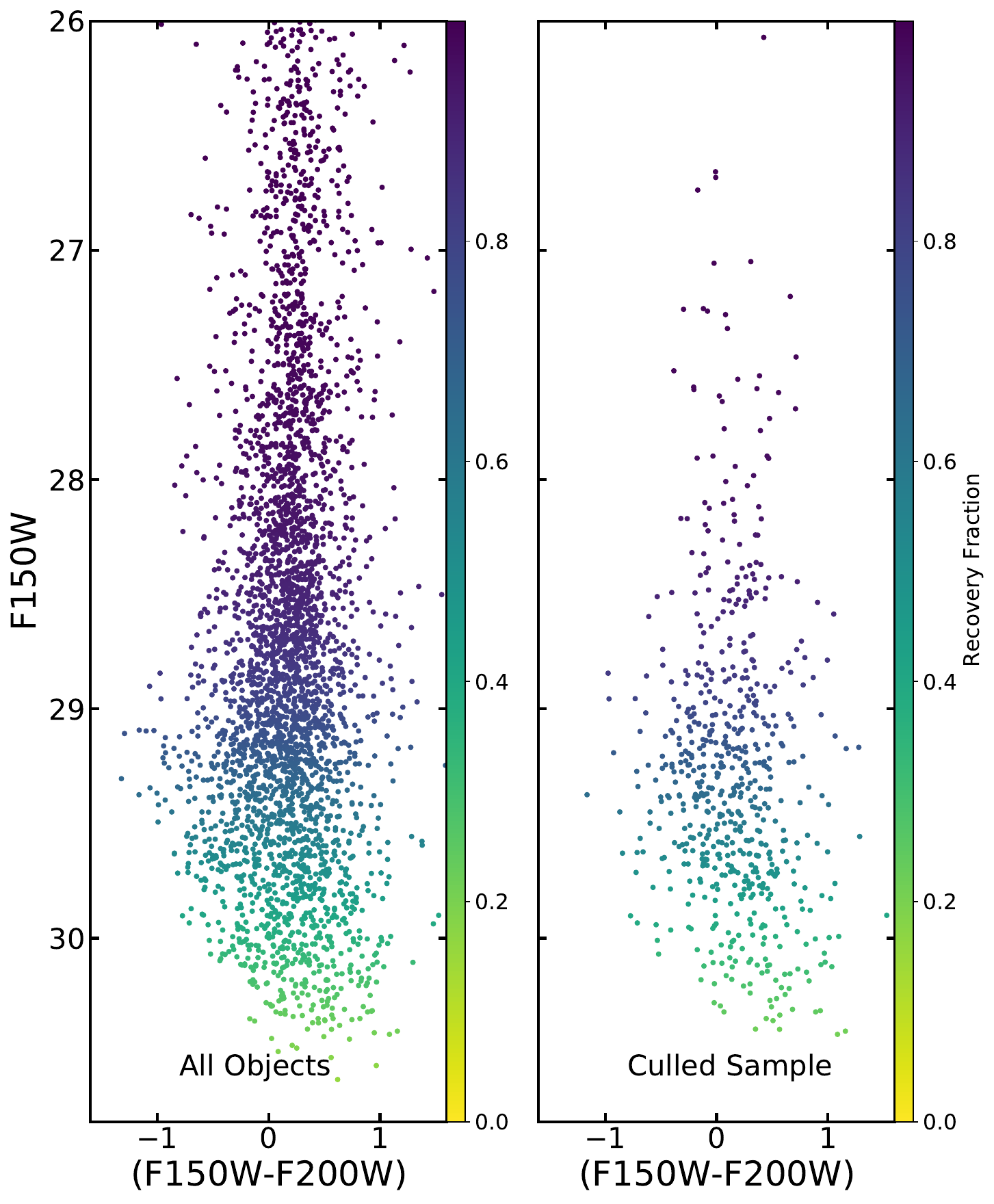}
  \caption{\emph{Left panel:} CMD in (F150W, F150W$-$F200W) for all 4508 measured objects in the field.  The majority of these are not point sources and therefore not candidates to be GC/UCDs.   The points are color-coded by the recovery fraction (completeness) indicated by the colorbsr.  \emph{Right panel:} CMD for the culled sample of 1009 distinguishable point sources. } 
  \label{fig:cmd2}}
\end{figure}

The effects of the culling are illustrated in Figures \ref{fig:xy2} and \ref{fig:cmd2}.  Figure~\ref{fig:xy2} (upper panel) plots every kind of object including small, faint background galaxies, nuclei of satellite galaxies, knots along lensing arcs or spiral galaxies, noise spikes and other artifacts, foreground stars, and the GC/UCD candidates.  Even so, distinct clumps of objects appear clustered around the BCG in the SE quadrant and the line of major galaxies in the NW quadrant. The application of \texttt{sharp} removes ${\sim}4/5$ of the objects, leaving the distribution as shown in  Figure~\ref{fig:xy2} (lower panel).  The concentrations around the major galaxies now show up more clearly, and the overall contamination level is much reduced.

Figure~\ref{fig:cmd2} shows the corresponding effects on the CMD\null, before and after removal of objects that are not point sources.  The distribution of the point sources across the El Gordo field is also marked in Fig.~~\ref{fig:field}, which shows that a high fraction of them are grouped around the major galaxies. For the BCG at lower left, it is notable that the point sources also follow the elongated shape of the BCG halo light. 

\section{Results}

\subsection{Isolating a GC/UCD Sample}

\begin{figure}
    \centering{
    \includegraphics[width=\columnwidth]{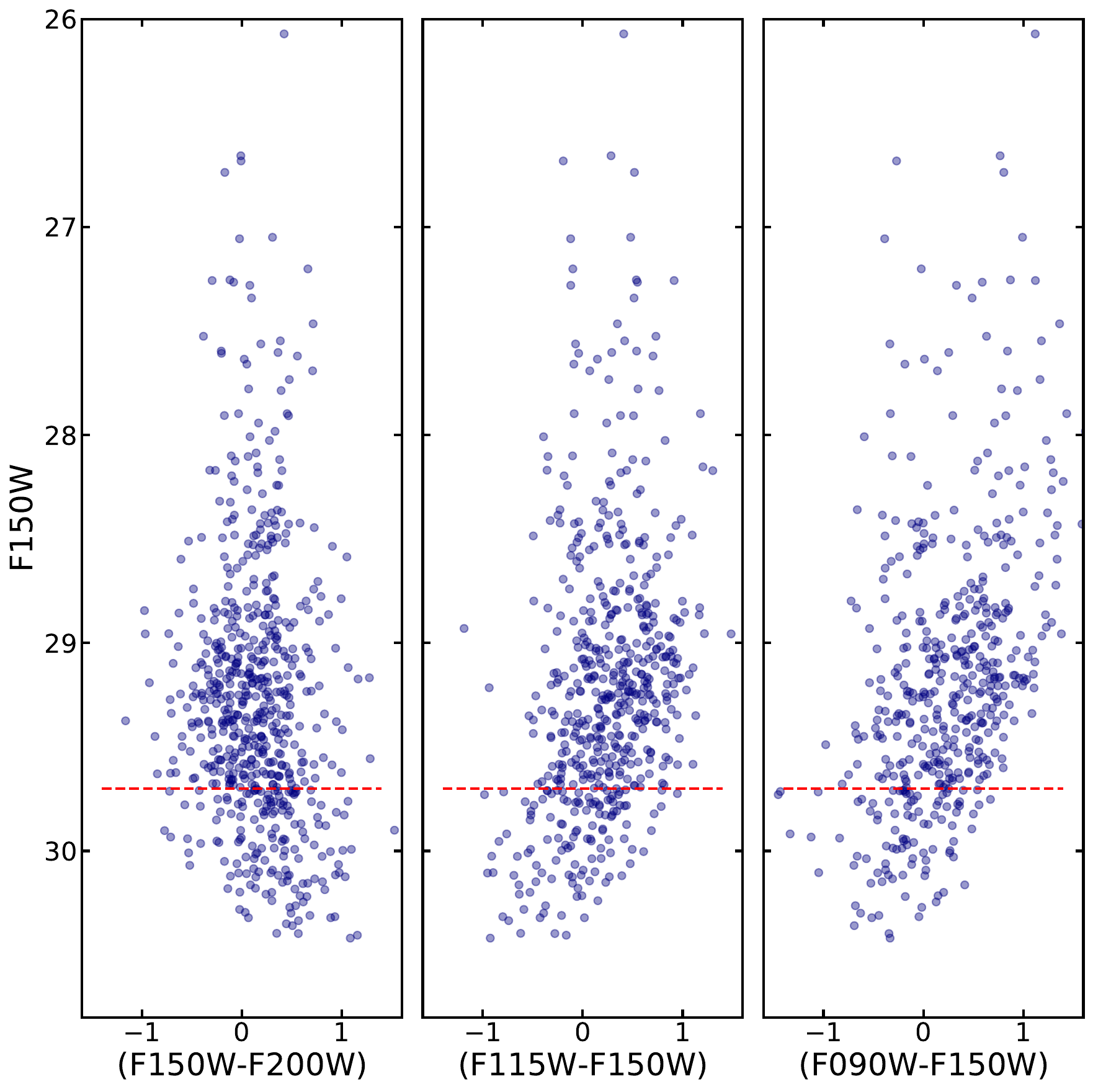}
  \caption{Color--magnitude diagrams for the culled sample of 1009 point sources in all quadrants. The F150W magnitude is plotted against three different color indices as labeled. Red dashed lines show the $p=0.5$ recovery limit.   Magnitudes and colors are as observed, i.e., no K-corrections have been applied.} 
  \label{fig:cmd3}}
\end{figure}

The final CMDs  for the culled point-source sample
are shown in Figure~\ref{fig:cmd3}. When all the point sources in the entire field are considered, no clear indication of a GC sequence is apparent.  However, when only the point sources \textbf{within the BCG region (defined as objects within the rectangle and circle in Figure~\ref{fig:xy2}) are examined,} a  sequence shows up, defined most narrowly in the (F150W-F200W) color. 

\begin{figure}
    \centering{
    \includegraphics[width=\columnwidth]{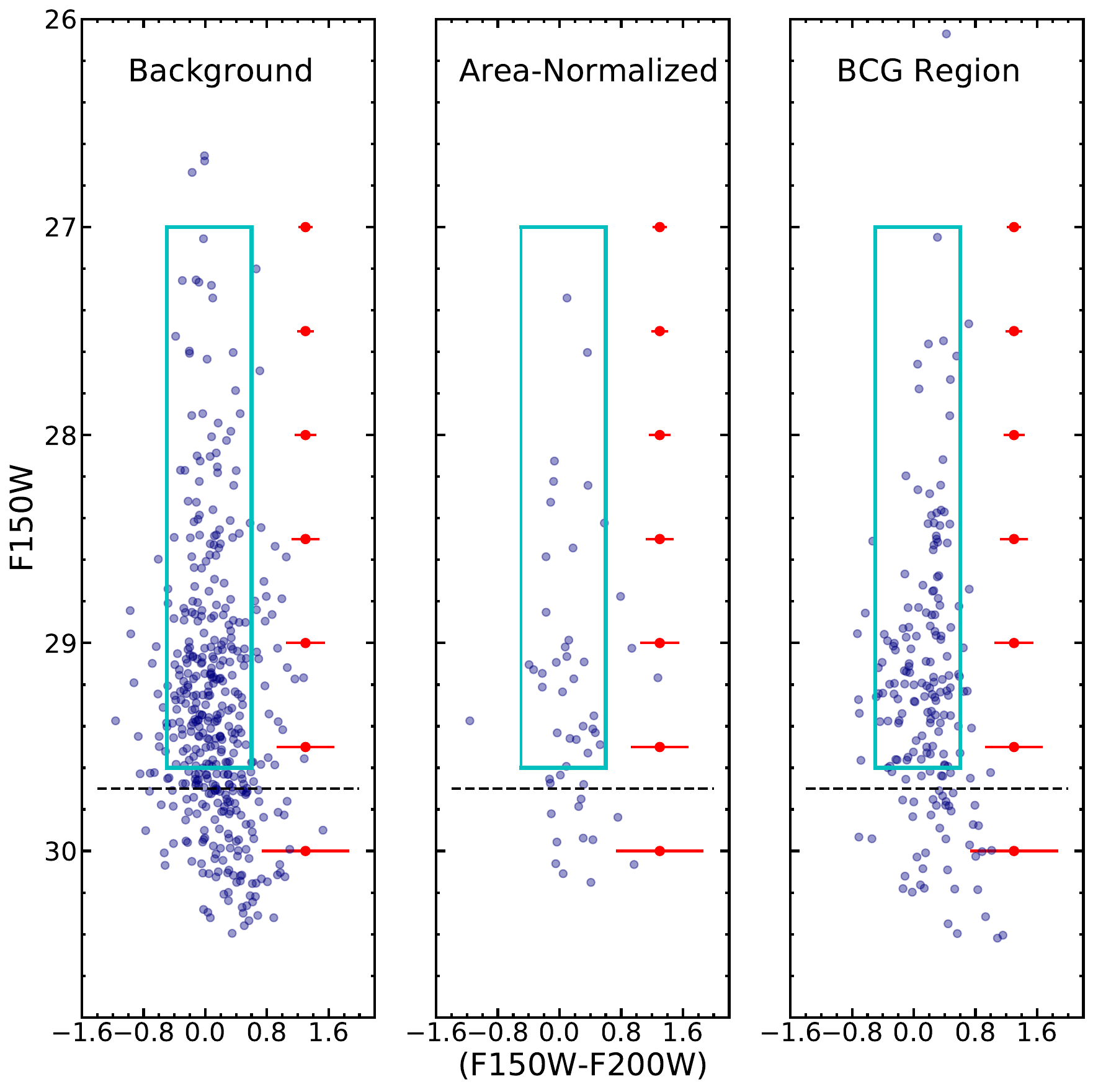}
  \caption{\textit{Left panel:} CMD for point sources outside the BCG region shown in 
  Figure~\ref{fig:xy2}.  \textit{Center panel:} same CMD as the left panel after randomly deleting 7/8 of the objects. This scales the background sample to the area of the BCG region.  \textit{Right panel:}  CMD for the point sources inside the BCG region.  In all panels, the \emph{cyan rectangles} 
  enclose the range $(-0.5, +0.6)$ in color and $(27.0, 29.6)$ in F150W\null. The red error bars to the right show the typical $1\sigma$ color uncertainties as a function of magnitude, and the horizontal dashed lines show the $p=0.5$ recovery limit.}  
  \label{fig:clean}}
\end{figure}

Figure~\ref{fig:clean} compares the CMDs for sources outside (left panel) and inside (right panel)  the BCG region.  The left panel defines what we call the ``background'' sample.  We narrow down the search further by using a region in the CMD defined by $-0.5 < \textrm{(F150W-F200W)} < +0.6$ and $27.0 < \textrm{F150W} < 29.6$ (cyan box in Fig.~\ref{fig:clean}).  This part of the CMD encloses the sequence matching the expected color range of the GC/UCDs and brighter than the limiting magnitude. The background region has 237 objects over an area of 10.7 arcmin$^2$, or 22 per arcmin$^2$.  The BCG region has 129 objects in an area of 1.35 arcmin$^2$ (1/8 the area of the background region), or 99 per arcmin$^2$. In the center panel of Fig.~\ref{fig:cmd3}, 1/8 of the objects in the background sample have been randomly selected to show the expected background population scaled to the same area as the BCG region.  This area-renormalized sample has 30 objects within the box. The excess of points within the BCG region is then $99\pm12$ objects, representing a highly significant detection of a GC/UCD population.

\subsection{Comparison with A2744}


At $z=0.31$, the A2744 system (HR23, HR24) is more evolved than the El Gordo system by 4 Gyr. Here we show a direct comparison of the CMDs for their GC/UCD populations, which requires applying both the difference in distance modulus and the K-corretions for the relevant filters.
The K-corrections are significant, but perhaps more importantly, they depend noticeably on GC metallicity and thus implicitly on age \citep{reina-campos_harris2024}.\footnote{A webtool to calculate K-corrections specifically for GCs is available at \url{https://rescuer.streamlit.app}.}  
In the Milky Way, the majority of GCs lie in the age range $\sim$9--13~Gyr following a rough age--metallicty relation \citep{leaman+2013,kruijssen+2019}.  At the lookback time of El Gordo, this range would be equivalent to 
2--6~Gyr, with the  metal-richest GCs expected to form systematically later.  The \emph{relative} age difference here is now more important: the higher-metallicity GCs will reach higher relative luminosity than the lower-metallicity ones because of the combined effects of stellar and dynamical evolution \citep{reina-campos+2022b}.  In addition, the higher-metallicity clusters are expected to form within higher-mass potential wells with more enriched gas, allowing them to reach higher masses than in the earlier, smaller, metal-poorer hosts \cite[e.g.,][]{li_gnedin2019,joschko+2024}.  At large lookback times, the classic `red' metal-rich GC sequence that appears in major galaxies should then extend to higher luminosity  than  the `blue' metal-poor sequence.  This difference is visible in several nearby BCG systems \citep{harris2023} but is not as prominent as it is expected to be  at larger lookback times.  Because younger GCs are more luminous than older ones, the bright end of the El Gordo GC sequence is likely to include primarily younger, metal-richer objects. 

Following the above argument, the metallicity for El Gordo clusters bright enough to be recovered in our data will tend to be high, so the K-corrections applied here are for solar metallicity. For an adopted mean age of 4~Gyr, the corrections are $K_{\rm F150W} = -0.05 \pm 0.06$~mag and $K_{\rm F200W} = -0.46 \pm 0.07$~mag.  The quoted uncertainties assume $\pm 0.3$~dex in metallicity and $\pm$20\% in age, but these K-values can differ by as much as $\simeq$0.2~mag over the known GC metallicity range $ [m/\rm H] \simeq -2$ to 0.  The effect on the color index $\rm (F150W-F200W)$ is similar, with $K_{(150-200)}$ ranging from $+0.2$ to $+0.4$~mag from low to high metallicity.

\begin{figure}
    \centering{
    \includegraphics[width=\columnwidth]{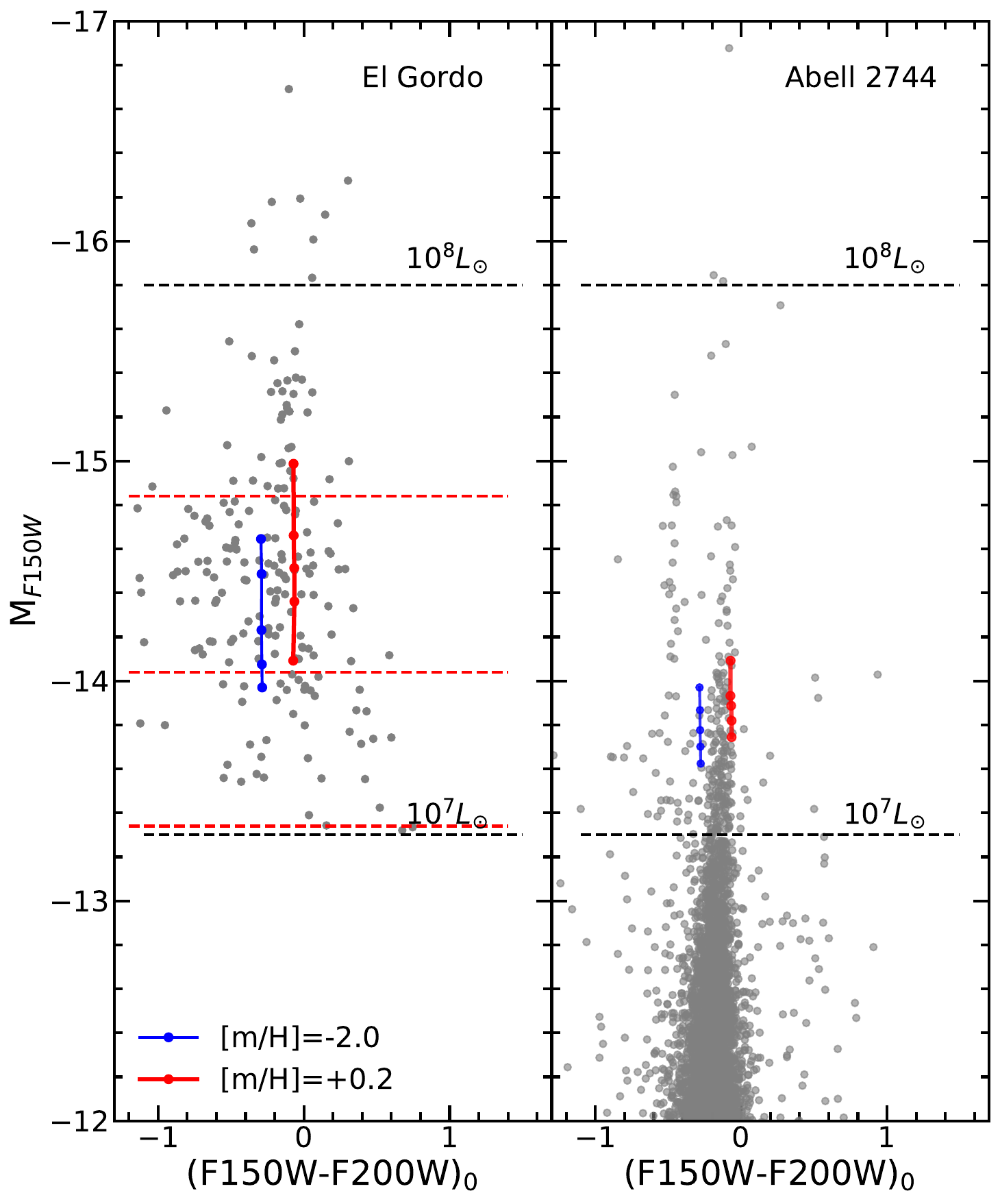}
  \caption{\textit{Left:} CMD for the GC candidates in El Gordo after cleaning.  Absolute magnitude is plotted against intrinsic color with K-corrections applied to both axes. (Note the axis scale change relative to the previous figure.) Completeness lines (top to bottom) for 80\%, 50\%, and 20\% are shown as red dashed lines. The {vertical lines and dots} indicate the predicted colors and magnitudes of a $5 \times 10^7$~\Msol\ GC at ages of 2, 3, 4, 5, and 6~Gyr from top to bottom. Blue lines and dots are for metallicity $[m/\rm H] =-2.0$, while the red ones are for  $[m/\rm H]  = +0.2$. The plotted colors and magnitudes are based on the PARSEC CMD3.7 simple stellar-population models \citep{bressan+2012}. \textit{Right:} Absolute magnitude versus intrinsic color for the A2744 GC population.  Here the blue and red lines and dots show the predicted locations of the same $5 \times 10^7 M_{\odot}$ objects but now at ages of 6, 7, 8, 9, and 10 Gyr to account for the 4-Gyr difference in lookback time between El Gordo and A2744. The 80\%, 50\%, and 30\% cpmpleteness limits lie below  the limit of the graph (HR23). Dashed horizontal lines in both panels indicate F150W luminosities as labeled. }
  \label{fig:cmdabs}}
\end{figure}

{Figure~\ref{fig:cmdabs} shows the CMDs for El Gordo and A2744 now converted to absolute magnitude and intrinsic color with the K-corrections applied.} For A2744 at much lower redshift, the K-corrections are less sensitive to metallicity or age; here we use $K_{150} = -0.16 \pm 0.03, K_{200} = -0.41 \pm 0.01$ mag for an adopted mean age of 8 Gyr.

For GCs several Gyr old, integrated color is insensitive to age (HR23), and the El Gordo GC sequence should have a mean intrinsic color similar to that of A2744. 
The mean color of the El Gordo GC sequence determined from the objects within $M_{\rm F150W} < -13.6$ and $-0.4 < \rm (F150W-F200W)_0 < +0.4$,  \textit{before} K-corrections, is $\langle \rm F150W-F200W\rangle = 0.20\pm 0.03$ with rms scatter  0.18~mag.  This  color would translate to a K-corrected value in the range $\langle\rm F150W-F200W\rangle_0 = -0.20$ to $-0.04$ depending on the actual mean metallicity along the sequence. By comparison, the A2744 GC sequence, where the K-corrections are smaller and better known, for the same color index over a similar luminosity range has a mean color of $-0.20$ with rms spread of 0.17 mag.\footnote{In the A2744 CMD, an `extra' blue sequence of points appears at $\rm (F150W-F200W) \simeq -0.5$.  These objects are scattered randomly around the field, and they do not define a clear sequence in other color indices, so they appear to be unassociated with the major galaxies (HR23).}

A more obvious difference between the El Gordo and A2744 CMDs appears in their luminosity (absolute magnitude) distributions, as shown in Figure~\ref{fig:cmdabs}.  The El Gordo GC sequence, though sparse, is more populated in the luminosity range $M_{\rm F150W} \la -14.7$ than A2744's GC sequence. This difference is unlikely to be due simply to a size-of-sample effect: in both clusters the measurements draw from GC populations around four or five massive early-type giant galaxies, and the virial masses of both clusters of more than  $10^{15} M_{\odot}$ are similar.  Because total GC population and the virial mass of the host cluster or galaxy are closely correlated \citep[e.g.,][]{blakeslee1997,spitler_forbes2009,harris+2017b,dornan_harris2025}, their GC populations should be similar. Instead, it is more likely that we are seeing the direct effect of luminosity evolution over the $\sim$4~Gyr difference in lookback time.  The dominant component over this time interval should be  straightforward stellar evolution. Mass loss due to tidal stripping and other dynamical losses will also occur but should be of minor importance at high mass:  for $M \simeq 10^7$~\Msol, the expected mass loss $\Delta M$ will correspond to  $\lesssim$0.1~mag of luminosity dimming \citep{li_gnedin2019}.  

The age-fading due to stellar evolution and its relation to age and metallicity are illustrated with model lines in Figure~\ref{fig:cmdabs}, as described in the figure caption.  
The decrease of luminosity with increasing age is  evident: the oldest, metal-poorest cluster in El Gordo is fainter than the youngest,  metal-richest one by 1~mag. Metallicity has a secondary effect, such that at any given age a higher-metallicity GC is slightly more luminous.  

For A2744, where the equivalent age range is 6 to 10~Gyr because of the 4-Gyr-smaller lookback time, GCs are fainter by $\simeq$0.8~mag compared with those in El Gordo. At this greater age, the \emph{relative} age difference between the youngest and oldest objects becomes less important. In A2744, the luminosity difference between the oldest, lowest-metallicity cluster and the youngest, highest-metallicity one is 0.47~mag, only half as much as in El Gordo.

As discussed in more detail by HR23, with only broadband photometry it is difficult to distinguish luminous GCs from their stellar-population cousins, the UCDs.  
In fact, the clearest part of the detected El Gordo `GC sequence' lies above $10^7$~\Lsol\ ($M_{\rm F150W}\simeq -13.3$).  This level has sometimes been used as a dividing line between GCs and UCDs, though there is no universally agreed-on transition level (see HR23 for references and discussion.)  This threshold, however, applies for systems \emph{at zero redshift}, and adjustment to the redshift of El Gordo would shift the threshold upward by a magnitude purely because of stellar evolution.  Taking this age effect into account, the point sources more luminous than $M_{150} \sim -14.3$ detected in El Gordo may therefore consist primarily of UCDs. 
We tentatively suggest that the small shelf of objects in the CMD starting at $\rm F150W \simeq 29$ ($M_{\rm F150W} \simeq -14.4)$ and $\rm (F150W-F200W) \simeq -0.5$ may represent the upper end of a true GC population.  Deeper, more precise photometry will be needed to reach any further conclusions.

A helpful observational feature distinguishing a true GC population would be the classic lognormal shape of its luminosity function (GCLF)  \citep[e.g.][]{jordan+2007,harris+2014}).  The EMP-Pathfinder simulations that trace the evolution of a GC system from its origin \citep{reina-campos+2022b} predict that the characteristic lognormal shape of at least the bright half of the GCLF is established within the first few Gyr.  Analysis of the A2744 GCLF (HR24) is consistent with this prediction, and indicates that the GCLF peak or turnover point lies at $M_{150} \simeq -8.7$.  However, the $\simeq$4~Gyr lower age of the El Gordo GCs would make the expected turnover point for El Gordo nearer $M_{150} \sim -8$, since the lower-mass GCs gradually get removed by dynamical destruction and the turnover shifts to higher luminosity with time.  This estimate would put the turnover point for the El Gordo GCs $\simeq$5~mag below the limit of the current photometry, far out of reach for the present.  Nevertheless, deeper exposures in future would allow at least a test of the shape of the GCLF over its brightest $\simeq$2--3 magnitudes, which would be a valuable test of the simulations.

\section{Summary and Future Prospects}

In this study we have used 
\jwst NIRCam imaging to detect and characterize GC/UCD candidates around the major galaxies in El Gordo at $z=0.87$.  Approximately 1--2 magnitudes of the GC/UCD sequence is clearly separated from field contamination.
The sequence begins at a luminosity level almost 1~mag brighter than in A2744 at $z\sim0.3$, consistent with the theoretical predictions about the decline of GC luminosity with age.

Strong limitations certainly apply to this initial look at the El Gordo system.  The measurements do not reach deeply into the GC/UCD sequence, and the metallicity distribution function cannot be measured because only the brightest candidates are detected, and the color index $\rm (F150W-F200W)$ is not intrinsically sensitive to metallicity for old GCs.  

 The work described here is reminiscent of the first detections of GCs around distant galaxies in the local Universe such as NGC~4874 (in Abell~1656, the Coma cluster), NGC~3842 (in Abell~1367), and NGC~6166 (in Abell~2199) \citep{harris1987,thompson_valdes1987,pritchet_harris1990,butterworth_harris1992}.  Only the brightest magnitude or a bit more of the GC luminosity functions was visible in these galaxies under the best ground-based observing conditions, but later and much deeper photometry with \hst confirmed these initial discoveries \citep{harris+2009,harris2023}.  That cycle was repeated with \hst imaging of the brightest GCs in much more distant targets such as Abell~1689 and Abell~2744, where again only the brightest magnitude or more of the GC/UCDs could be resolved  \citep{lee_jang2016,alamo-martinez+2013,alamo-martinez+2017}.  Again, much deeper photometry with JWST has confirmed these detections.  In both cases the  available imaging tools of the day were being pressed to their limits.  For the present NIRCam imaging of El Gordo, the existing \jwst exposures are fairly short, but they are deep enough to  identify the bright end of the GC/UCD sequence.  There is certainly room for deeper probing into the system.

Further study of GC/UCD populations in host clusters of galaxies at redshifts approaching one will require deeper observations.
Nevertheless, the present data, despite their modest depth, demonstrate that detection and measurement of the GC/UCD populations in large galaxies at $z \sim 1$ is readily possible with \jwst.  The results also provide an empirical  gauge of the limit to which GC populations can be directly imaged by \jwst.

Successful photometry of GC populations around distant galaxies depends on \textit{photometric depth, spatial resolution,} and \textit{sample contamination:}
\begin{enumerate}
    \item Depth (limiting magnitude) is determined by total exposure time. Photometric limits $m_{\rm AB} \gtrsim 30$ have been achieved in other studies (HR24) and will allow detection of the brightest GCs at $z > 1$.
    \item Spatial resolution determines the degree of crowding of the GCs that can be tolerated.  For El Gordo, crowding proved not to be an issue except for the innermost $\sim$10\arcsec\ around the major galaxies.  At larger distances, the level of crowding will be little different because the angular size distance $d_A = d_L/(1+z)^2$ changes very slowly with redshift \citep[e.g.,][]{condon_matthews2018}. 
    \item A more important factor is field contamination:  if, for example, the El Gordo GC/UCD candidates were one magnitude fainter (Figure~\ref{fig:clean}), field contamination would begin to dominate the measured point sources, and it would be much more difficult to extract the GC sequence.  Taking into account the higher GC luminosity at larger lookback times, an empirical estimate of the practical distance limit is then $(m-M)\sim 45$ or $d_L \sim 10$~Gpc and $z_{\rm lim} \simeq 1.4$.  At this distance, the lookback time is 9~Gyr, and classic GCs and UCDs would be  just 1--4~Gyr old.  That limit may be too conservative, however, because longer exposure times and more precise photometry will allow  more stringent removal of nonstellar objects and thus  reduction of field contamination. 
\end{enumerate}

GC photometry in target galaxies at $z \sim 1.0$--1.5  will  open a window into the state of GC systems at times approaching `cosmic noon', the epoch of maximum star-formation rate in galaxies \citep[e.g.,][]{madau_dickinson2014}.  In addition, $z_{\rm lim} \gtrsim 1.4$  for `normal' imaging of GC systems comes close to overlapping the redshift range at which candidate young GCs have been detected by strong lensing \citep{mowla+2022,adamo+2023,claeyssens+2023,vanzella+2022,messa+2024}. 
Ultimately, the prospects are encouraging for eventually seeing the entire history of GC evolution by direct observation.



\begin{acknowledgments}

MRC gratefully acknowledges the Canadian Institute for Theoretical Astrophysics (CITA) Fellowship for support. This work was also supported by the Natural Sciences and Engineering Research Council of Canada (NSERC). RAW acknowledges support from NASA JWST Interdisciplinary Scientist grants NAG5-12460, NNX14AN10G and 80NSSC18K0200 from GSFC.

\end{acknowledgments}

The data used in this paper were obtained from the Mikulski Archive for Space Telescope (MAST) at the Space Telescope Science Institute.  The specific observations analyzed can be accessed via \dataset[doi: 10.17909/x49n-d207]{https://doi.org/10.17909/x49n-d207}.
\facilities{\jwst (NIRCAM)}


\software{Daophot \citep{stetson1987},
          IRAF \citep{tody1986,tody1993}
          Jupyter Notebooks \citep{kluyver+2016}, 
          Matplotlib \citep{hunter2007},
          Numpy \citep{harris+2020b},
          PARSECv1.2 \citep{bressan+2012}
          }
          
\bibliography{main}{}
\bibliographystyle{aasjournalv7}



\end{document}